\documentclass[twocolumn,prd,superscriptaddress,showpacs,floatfix,  
preprintnumbers, nofootinbib,hyperref]{revtex4} 
\usepackage{epsfig}   
\usepackage{bm}  
\usepackage{color}  
\usepackage{amsmath}

\newcommand{\ben}{\begin{equation}}
\newcommand{\een}{\end{equation}}

\newcommand{\gtrsim}{\,\rlap{\lower3.7pt\hbox{$\mathchar\sim$}}
\raise1pt\hbox{$>$}\,}
\newcommand{\lesssim}{\,\rlap{\lower3.7pt\hbox{$\mathchar\sim$}}
\raise1pt\hbox{$<$}\,}

\newcommand{\be}{\begin{equation}}
\newcommand{\ee}{\end{equation}}  
\newcommand{\bea}{\begin{eqnarray}}
\newcommand{\eea}{\end{eqnarray}}

\begin{document}

\title{Analysis of matter suppression in collective  neutrino oscillations \\
during  the supernova accretion phase}

\author{Sovan Chakraborty} 
\affiliation{II Institut f\"ur Theoretische Physik, Universit\"at Hamburg, Luruper Chaussee 149, 22761 Hamburg, Germany}
 
\author{Tobias Fischer} 
\affiliation{GSI, Helmholtzzentrum f\"ur Schwerionenforschung GmbH,
Planckstra{\ss}e 1
64291 Darmstadt, Germany}
\affiliation{Technische Universit\"at Darmstadt, Schlossgartenstra{\ss}e 9, 64289 Darmstadt,
Germany}

\author{Alessandro Mirizzi} 
\affiliation{II Institut f\"ur Theoretische Physik, Universit\"at Hamburg, Luruper Chaussee 149, 22761 Hamburg, Germany}
 
\author{Ninetta Saviano} 
\affiliation{II Institut f\"ur Theoretische Physik, Universit\"at Hamburg, Luruper Chaussee 149, 22761 Hamburg, Germany} 
 
\author{Ricard Tom{\`a}s}   
\affiliation{II Institut f\"ur Theoretische Physik, Universit\"at Hamburg, Luruper Chaussee 149, 22761 Hamburg, Germany}

\begin{abstract}
The usual description of self-induced neutrino flavor conversions in core collapse
supernovae (SNe) is based on the dominance of the neutrino density $n_{\nu}$
over the net electron density $n_{e}$. However, 
this condition is not met during the post-bounce accretion
phase, when the dense matter in a SN is piled up above the neutrinosphere.
As recently pointed-out, a dominant matter term 
in the anisotropic SN environment would dephase the flavor evolution for
neutrinos traveling on different trajectories, challenging the occurrence of
the collective behavior in the dense neutrino gas. 
Using the results from recent long term simulations of core-collapse SN
explosions, based on three flavor Boltzmann neutrino transport in spherical
symmetry, we find that both the situations of complete matter suppression
(when $n_{e}\gg n_{\nu}$) and matter-induced decoherence
(when $n_{e}\gtrsim n_{\nu}$) of flavor conversions are realized during the accretion phase. 
The matter suppression at high densities prevents any possible impact of the
neutrino oscillations on the neutrino heating and hence on the dynamics of
the explosion.
Furthermore, it  changes the interpretation
of the Earth matter effect on the SN neutrino signal during the accretion phase,
allowing the possibility of  the neutrino mass hierarchy discrimination at not too small
values of the leptonic mixing angle $\theta_{13}$
(i.e. $\sin^2{\theta}_{13} \gtrsim 10^{-3}$) .  
\end{abstract}


\maketitle

\section{Introduction} 

Massive star explosions are an active subject of research,
in terms of core-collapse supernova (SN) models.
The supernova problem is related to the revival of the stalled bounce
shock, which forms when the collapsing stellar core reaches nuclear matter density
and bounces back.
It propagates out of the core and stalls on its way out due to the continuous
energy loses via neutrino emission
and dissociation of heavy nuclei.
Independently from the explosion mechanism (see, 
e.g.~\cite{LeBlancWilson:1970,BetheWilson:1985,Burrows:2006,Kitaura:2006,Fischer:2009af,
Sagert:2009,Fischer:2010})  the total energy emitted in neutrinos
($\nu$) and antineutrinos ($\bar{\nu}$) during a SN is of the order of several
$10^{53}$~erg, which makes a SN the most powerful neutrino source in the
Universe.
The total duration of the neutrino emission, known as the neutrino burst,  can last up
to 30~seconds.
This $\nu$ signal represents a powerful
tool to probe fundamental neutrino properties as well as the dynamics of the
explosion~\cite{Dasgupta:2010gr}.

The role of astrophysical messengers played by neutrinos during a stellar
collapse is largely associated with the signatures imprinted on the observable
SN neutrino burst by  weak processes and  flavor conversions
occurring deep inside the star.
In the last few years, it has been understood that the description of neutrino
flavor conversions in supernovae, based on the only
Mikheyev-Smirnov-Wolfenstein~(MSW)  effect with the ordinary
matter~\cite{Matt}, was incomplete.
In particular, in the deepest supernova regions  
the neutrino density is so high that the neutrino-neutrino interactions dominate
the flavor changes.
Even if the important role played by the neutrino-neutrino interactions for the flavor
evolution in supernovae was pointed out since a long time
ago~\cite{
Pantaleone:1994ns,
Qian:1993dg,
Qian:1994wh,
Pastor:2002we,
Fuller:2005ae},
 only recently the first numerical experiments have been performed
to characterize realistically these effects~\cite{Duan:2005cp,Duan:2006an}.
These seminal investigations have stimulated a still streaming torrent of
activities (see, e.g.~\cite{Duan:2010bg}, for a recent review).
The general result of these studies is that neutrino-neutrino interactions
create a large potential for the neutrinos, which causes large
and rapid conversions between different flavors.
The transitions occur collectively, i.e. in a coherent fashion,
over the entire energy range. 
The most important observational consequence of $\nu$--$\nu$ interactions is a
swap of the $\nu_e$ and ${\overline\nu}_e$ spectra with the non-electron $\nu_x$
and ${\overline \nu}_x$ spectra in certain energy
ranges~\cite{
Duan:2006an,Duan:2010bg,
Fogli:2007bk,
Fogli:2008pt,
Raffelt:2007cb, 
Raffelt:2007xt,
Duan:2007bt,
Duan:2008za,
Gava:2008rp,
Gava:2009pj,
Dasgupta:2009mg,
Fogli:2009rd,
Chakraborty:2009ej,
Friedland:2010sc,
Dasgupta:2010cd,
Duan:2010bf,
Mirizzi:2010uz,
Galais:2011jh,Galais:2011gh}.

The basic idea behind the self-induced neutrino oscillations in supernovae
is that these can produce significant flavor
conversions close to the $\nu$ decoupling region (i.e., at small radii)  even in the presence
of a large matter density~\cite{Duan:2005cp,Hannestad:2006nj}. This
 would have inhibited any possible flavor transition, far from the MSW resonances
(i.e., far outside the neutrino-matter decoupling region) at larger radii~\cite{Dighe:1999bi}. 
Subsequently, it has been realized in~\cite{EstebanPretel:2008ni}, and then confirmed
in~\cite{Duan:2008fd},
that this original idea was only in part true since the matter density cannot be arbitrarily
large before it affects collective flavor conversions after all.
Indeed, for the strongly anisotropic emission of neutrinos streaming-off the SN core,
the matter effect is trajectory dependent, inducing different oscillation phases for
neutrinos traveling in different directions.
When the neutrino-neutrino interaction is sufficiently strong, it forces  neutrinos reaching
the same position from different paths  to have  the same oscillation phase, overcoming
the trajectory-dependent dispersion induced by the matter.
Conversely, when the electron density $n_e$ significantly exceeds the neutrino
density $n_\nu$, it has been shown that the large phase dispersion induced by the
matter would suppress the collective phenomena.
Depending on the electron density, the matter suppression can be
complete, when $n_e\gg n_\nu$, or partial when the matter
 domination  is less pronounced.
Finally, when $n_e \gtrsim n_\nu$,  the interference of the
two comparable effects would lead to the decoherence of the collective neutrino
flavor changes, producing an equal mixture between the 
oscillating electron and non-electron neutrino species~\cite{EstebanPretel:2008ni}.

Most of the numerical simulations of the self-induced flavor conversions in SN
explosions  have been focused on the late times cooling phase
(at post-bounce times $t_{pb} \gtrsim 1$~s), when
the matter potential is smaller than the neutrino-neutrino potential. 
In such a situation, collective oscillations would develop without any matter
hindrance~\cite{Duan:2006an,Fogli:2007bk,Duan:2010bf,Mirizzi:2010uz}. 
During the earlier phase before the onset of an explosion, which is 
 determined by mass accretion ($t_{pb} \lesssim 0.5$ s), the electron density  is not
negligible with respect to the neutrino  density, as one can 
infer from the following back--of--the--envelope calculation.
Given the mass accretion rate 
\begin{equation}
\dot{m} = \frac{\partial m}{\partial r} v = 4\pi\rho r^2 v \,\ ,
\end{equation}
in terms of the the rest-mass density $\rho$   at radius $r$, and of the matter velocity $v$,
one gets 
\begin{equation}
\rho \sim 10^8\,\frac{\textrm{g}}{\textrm{cm}^3} \left(\frac{\dot{m}}
{0.4\textrm{ M}_{\odot}{/s}} \right)
\left(\frac{10^2\textrm{ km}}{r}\right)^2 
\left(\frac{10^4\textrm{ km}/\textrm{s}}{v} \right) ,
\end{equation}
for typical values in the region ahead the shock-wave.
Then, the net electron density at $r \sim 10^2$~km can be estimated as  
\begin{equation}
 n_e = \rho Y_e/m_B \sim 10^{32} \,\ \textrm{cm}^{-3} \,\ ,
\end{equation}
where the electron fraction  $Y_e \simeq 0.5$ and $m_B$ is the nucleon mass. 
On the other hand, 
during the accretion phase, where the charged-current reactions dominate,
the  far-distance  electron (anti)neutrino luminosities can be
expressed via the change of the gravitational potential at the neutrinosphere
\footnote{The neutrinospheres are the neutrino energy and flavor dependent
spheres of last scattering.}
with radius $r_\nu$ as follows,
\begin{equation*}
L_{\nu_{e},\bar{\nu}_{e}} = \left . \frac{Gm}{r}\,\dot{m}\right\vert_{r_{\nu_e,\bar\nu_e}}
\end{equation*}
\begin{equation}
\sim
10^{52}\,\frac{\textrm{erg}}{\textrm{s}} 
\left(\frac{{m}}{1.5\textrm{ M}_{\odot}} \right)
\left(\frac{\dot{m}}{0.4\textrm{ M}_{\odot}{/s}} \right)
\left(\frac{10^2\textrm{ km}}{r}\right).
\end{equation}
Therefore,  the  neutrino particle flux
$F_{\nu_{e},\bar{\nu}_{e}} =
L_{\nu_{e},\bar{\nu}_{e}} /\langle E_{\nu_{e},\bar{\nu}_{e}} \rangle
\sim 10^{57} $~s$^{-1}$,
for $\langle E_{\nu_{e},\bar{\nu}_{e}} \rangle= 11$~MeV, implying
\begin{equation}
n_{\nu_{e},\bar{\nu}_{e}} = \frac{F_{\nu_{e},\bar{\nu}_{e}}}{4 \pi r^2}
\sim 10^{31}{\textrm{cm}^{-3}} \,\ ,
\label{eq:nv}
\end{equation}
at $r \sim 10^2$~km.
From this simple estimation, $n_\nu$  can be smaller or on the same order
than  $n_e$ and hence we cannot a-priori assume
that $n_e \ll n_{\nu}$ holds during the post-bounce accretion phase.
This result is confirmed by many different SN
simulations~\cite{shock,Tomas:2004gr,Buras:2005rp, Liebendoerfer:2003es}.
It is a generic feature that applies generally to core collapse SNe of massive iron-core
progenitors.
As an important consequence, it is not guaranteed a priori that the results of the
neutrino flavor evolution obtained during the cooling phase, could be directly applied
to the accretion.

The post-bounce accretion phase, where the neutrino fluxes are large, represents
a particularly interesting scenario for detecting signatures of neutrino flavor 
transformations.
Indeed, also the flavor-dependent flux differences are large with a robust
hierarchy for the neutrino number fluxes,
$F_{\nu_e} > F_{{\overline\nu}_e} \gg F_{\nu_x}$,
where $\nu_x$ indicates the non-electron flavors.
In such a situation, if dense matter effects are neglected and the flavor asymmetries
are enough to prevent multi-angle effects~\cite{EstebanPretel:2007ec}, the
self-induced flavor oscillations seemed to have a clear outcome, producing
a complete exchange of the electron and non-electron flavors for almost
all antineutrinos in the inverted  neutrino mass hierarchy
(IH, $\Delta m^2_{\rm atm} = m_3^2 - m_{1,2}^2<0 $) 
and a spectral split in the energy distributions of the neutrinos~\cite{Fogli:2007bk}.
In the normal neutrino mass hierarchy (NH, $\Delta m^2_{\rm atm} > 0$),
they would leave the neutrino spectra unaffected.
How this picture would change if the  effect of the dense matter is included,
is subject of investigation of the present article. It develops and
 clarifies   the results recently presented
in our~\cite{Chakraborty:2011nf}.

We take as benchmark the results of the recent long-term
SN simulations from Ref.~\cite{Fischer:2009af}, to characterize the SN neutrino signal
as well as the matter density profile evolution. 
We consider three cases corresponding to different supernova progenitor masses, 
the 8.8~M$_\odot$ O-Ne-Mg-core and the 10.8~M$_{\odot}$ and 18.0~M$_{\odot}$
iron-cores. 
For these three cases we find that the matter density is always larger than or
comparable to the neutrino density, during the accretion phase. 
This finding suggests the possible hindrance (or decoherence) of collective
neutrino flavor transitions for matter densities higher than (or comparable with)
the neutrino densities during the accretion phase.

The plan of our paper is the following.
In Section~II we describe the original supernova neutrino emission during the accretion
phase and in Section~III we introduce  the schematic supernova neutrino model we use
to characterize the $\nu$ emission geometry, the neutrino potentials and the different
radial ranges where self-induced oscillations and matter
effects  are  relevant.
The results of our analysis on matter suppression for the three different SN progenitor
masses are presented in Section~IV and the impact of the dense matter suppression 
on the interpretation of the Earth matter effect on the SN neutrino burst
is discussed in Section~V. 
Finally, we close with comments of our results and the conclude
in Section~VI.
 
\begin{figure}  
\includegraphics[angle=0,width=.49\textwidth]{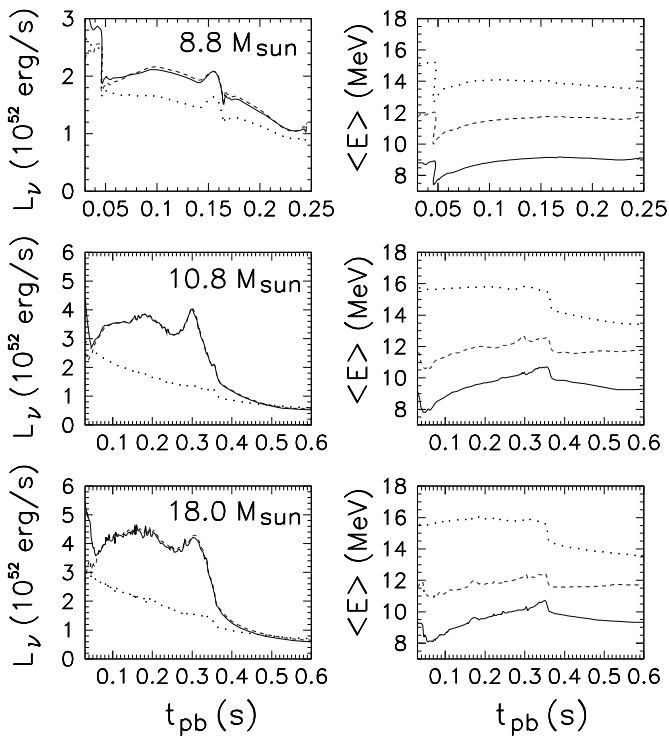}  
\caption{
Neutrino luminosities (left panels) and average energies (right panels)
for the core-collapse SN simulations considered, where we distinguish
$\nu_e$ (continuous line), $\bar\nu_e$ (dashed line) and $\nu_x$ (dotted line). 
The three cases are shown   corresponding to different values of the supernova
progenitor mass: 8.8~M$_\odot$ (top), 10.8~M$_\odot$ (middle) and
18.0~M$_\odot$ (bottom). 
\label{fig1}}  
\end{figure}  
 
\section{Neutrino signal from core-collapse supernova explosions}

We investigate core-collapse supernova simulations
of massive progenitor stars of 8.8, 10.8 and 18~M$_\odot$.
The first one belongs to the class of O-Ne-Mg-core progenitors
\cite{Nomoto:1983,Nomoto:1987}
and represents the threshold between thermonuclear
explosions and core-collapse supernovae \cite{Kitaura:2006, Fischer:2009af}.
The latter two are iron-core progenitors \cite{Woosley:2002}.
All models were evolved consistently through core collapse,
bounce and the early post-bounce phase up to several seconds
after the onset of explosion~\cite{Fischer:2009af}.
The core-collapse model is based on general relativistic
radiation hydrodynamics that employs three flavor Boltzmann neutrino
transport in spherical symmetry and a sophisticated equation of state
for hot and dense nuclear matter~\cite{Shen:1998}
(for details about the supernova model, see~\cite{Liebendoerfer:2004,Fischer:2009af}
and references therein).
Explosions of massive iron-core progenitors cannot
be obtained in spherically symmetric supernova models.
In order to trigger the explosions for the 10.8 and 18~M$_\odot$ progenitor
models, the heating rates are artificially enhanced in the gain region where
neutrinos deposit energy in order to revive the stalled bounce shock.

Fig.~\ref{fig1} shows the evolution of the neutrino luminosities
$L_{\nu_\alpha}$ (left panels) and average energies $\langle E_{\nu_\alpha}\rangle$
(right panels), during the post-bounce accretion phase
for all models under investigation.
Here, $\nu_\alpha = (\nu_e,{\overline\nu}_e,\nu_x)$ where
$\nu_x$ indicates both $(\mu,\tau)$-neutrinos and antineutrinos.
The large electron flavor neutrino luminosities $\mathcal{O}(10^{52})$~erg/s are
 due to the continuous mass accretion at the neutrinospheres, where the
electron flavor neutrino luminosities are dominated by the charged current reactions.
The slightly larger ${\overline\nu}_e$ luminosity over the $\nu_e$ luminosity and
the magnitude of the differences obtained in the models discussed here, is an
active subject of research and may change slightly applying improved weak rates
and multi-dimensional supernova models.
The increasing and decreasing electron flavor luminosities reflect the propagating
bounce shock (and the mass accretion rate at the neutrinospheres), which contracts
and expands accordingly during the accretion phase driven by neutrino heating/cooling.
Since muonic charged current processes are suppressed,  $\nu_{\mu/\tau}$'s are
produced only after bounce via pair-processes and the $(\mu/\tau)$-neutrino
luminosities are generally smaller than the electron flavor neutrino luminosities.
For the average energies, the following hierarchy holds
$\langle E_{\nu_e}\rangle < \langle E_{\bar{\nu}_e}\rangle < \langle E_{\nu_x}\rangle$
for all models under investigation during the accretion phase.
Furthermore, the average energies rise continuously
for all neutrino flavors during  the accretion phase illustrated
in Fig.~\ref{fig1} (right panels).

After the onset of explosion,  when mass accretion vanishes,
the situation changes.
The luminosities and average energies of all flavors decrease
on a longer timescale on the order of seconds.
For the O-Ne-Mg-core, we consider the neutrino signal only up
to $t_{\rm pb}$=0.25~s (top panel of Fig.~\ref{fig1}).
Note that mass accretion at the neutrinospheres vanishes already at about
$t_{\rm pb}$=0.03~s.
This determines the onset of the explosion for this low-mass progenitor.
For the 10.8 and 18~M$_\odot$ progenitor models discussed here, we show
the neutrino signal up to $t_{\rm pb}$=0.6~s.
For these cases, the onsets of explosion occur at about $t_{\rm pb}$=0.36~s,
due to the more massive envelopes surrounding the iron-core that
lead to a more extended accretion (and hence neutrino heating) phase.
Note the sharp drop of the luminosities and average energies of all neutrino flavors
in Fig.~\ref{fig1} after the onsets of explosion.
These are due to the sudden flip of matter velocities from infall to expansion when
the explosion shock passes through the distance of 500~km, where the observables
are measured in a co-moving frame of reference.
We note further that these results suggest lower average energies
than often assumed in the literature~\cite{Totani:1997vj} and a less pronounced
spectral hierarchy, in particular during the later proto-neutron star cooling phase.
The results obtained for the low mass O-Ne-Mg-core collapse supernova explosion
and long term simulation of the proto-neutron star cooling phase, are in qualitative
and quantitative good agreement with recent simulations performed by the
Garching-group~\cite{Huedepohl:2009}. 

\section{Schematic supernova model} 

\begin{figure}  
\includegraphics[angle=0,width=0.49\textwidth]{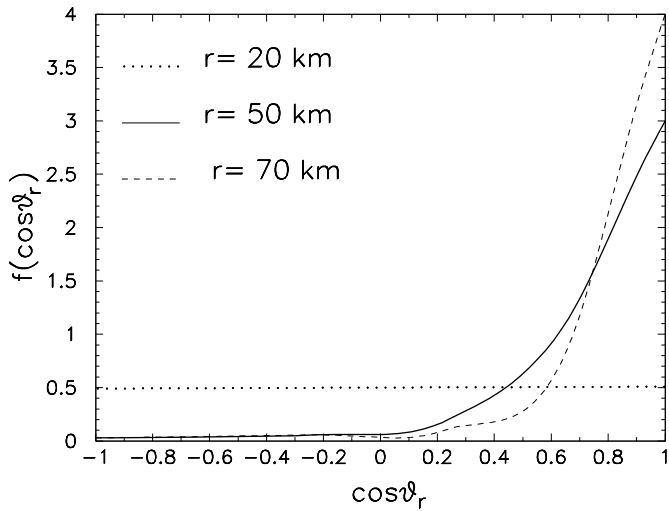}  
\caption{10.8 M$_{\odot}$ progenitor mass. Angular distributions for $\nu_e$ at $t_{\rm pb}$=0.4~s for $E_{\rm rms}=11.5$~MeV
 at different radii: r=20~km (dotted line),   r=50~km (continuous line), 
  r=70~km (dashed  line). 
\label{fig2}}
\end{figure}  

We present here our schematic assumptions to characterize the flavor conversions 
during the accretion phase.
In particular, we describe our model for the SN neutrino emission geometry, 
the neutrino number fluxes of different flavors, 
the neutrino-neutrino and matter potentials and the different oscillation regimes.

\subsection{Neutrino emission geometry} 

We assume a spherically symmetrical source (``SN core'') that emits neutrinos and
antineutrinos like a blackbody surface, from a neutrinosphere at radius $r=r_\nu$. 
We define $r_\nu$ to be the radius where \color{black} the neutrino radiation
field is assumed to be ``half-isotropic'', i.e. all outward-moving angular modes are
equally occupied~\cite{EstebanPretel:2007ec}.
We remind that this conventional definition of the neutrinosphere, used in the context
of the neutrino flavor conversions, is  only intended  to fix a boundary
condition for the subsequent flavor evolution.
We assume that weak processes take place at higher densities inside the
neutrino-matter decoupling sphere and are negligible outside the decoupling sphere
where neutrino oscillations can occur.
The half-isotropic definition for the decoupling spheres  does not
necessarily coincide with  the definition of the neutrinosphere as neutrino last
scattering surface,  defined as the radius where the optical depth becomes
2/3.
However, in the context of collective oscillations which start at radii
much larger than the assumed boundary, neutrinos are safely in a free-streaming
regime~\cite{Duan:2010bf}. Therefore, we do not have to worry about the details
of the $\nu$ decoupling.

Instead of choosing arbitrarily the neutrinosphere radius, we fix it consistently with
the  SN simulations which provide the angular distributions of the different neutrino 
species as a function of time and energy at different radii.
For definiteness, we have taken the $\nu_e$'s  distribution as representative 
for all the different flavors, since electron neutrinos reach the free-streaming regime
at larger radii compared to the other neutrino species during the accretion
phase. The angular distributions are functions of the energy. For definiteness, 
we consider as reference  the root mean square (rms) energy relevant
for the average neutrino-nucleon interaction rates, that determine the neutrino
angular distributions.
As an example, in  Fig.~\ref{fig2} we plot the $\nu_e$ angular distribution 
$f(\cos \theta_r) = d n_\nu/d \cos \theta_r$, where $\theta_r$ is the zenith angle of  a given
mode relative to the radial direction at distance $r$.  
We consider 
$E_{\rm rms}= 11.5$~MeV  at three different radii at the post-bounce time
$t_{\rm pb}=0.4$~s, taken form the 10.8~M$_{\odot}$ model.
As expected, at  small  radii ($r=20$~km in the Fig.~\ref{fig2})
the angular distribution is isotropic since neutrinos are in a trapping regime.
They are isotropically emitted in all directions.
At large radii ($r=70$~km in the Fig.~\ref{fig2}), neutrinos are free-streaming and
their angular distribution becomes forward peaked.
We schematically \emph{assume} as neutrinosphere radius, the one at which the
$\nu_e$'s angular distribution has no longer significant backward flux, i.e. a few $\%$ of
the total one  (at $r=50$~km in the Fig.~\ref{fig2}).
Even if there the real angular distribution is not half-isotropic, we assume
it to characterize the further flavor evolution. For our purpose of
evaluating the impact of the matter effects on the self-induced flavor
conversions,  this simplified choice is conservative. Indeed,
the real angular distributions, which  are more  forward-peaked than what we are assuming, 
would reduce the real strength of the neutrino-neutrino potential with respect to what
we are assuming. 

We stress that this procedure for fixing the neutrinosphere has to be taken only as an
empirical prescription. 
However,  we performed several neutrino flavor oscillation analysis
changing by a factor of three the neutrinosphere radius. We find that
the results on matter effects remained basically unchanged, 
indicating a weak dependence of the oscillation analysis from the
decoupling region.

\begin{figure}  
\includegraphics[angle=0,width=0.49\textwidth]{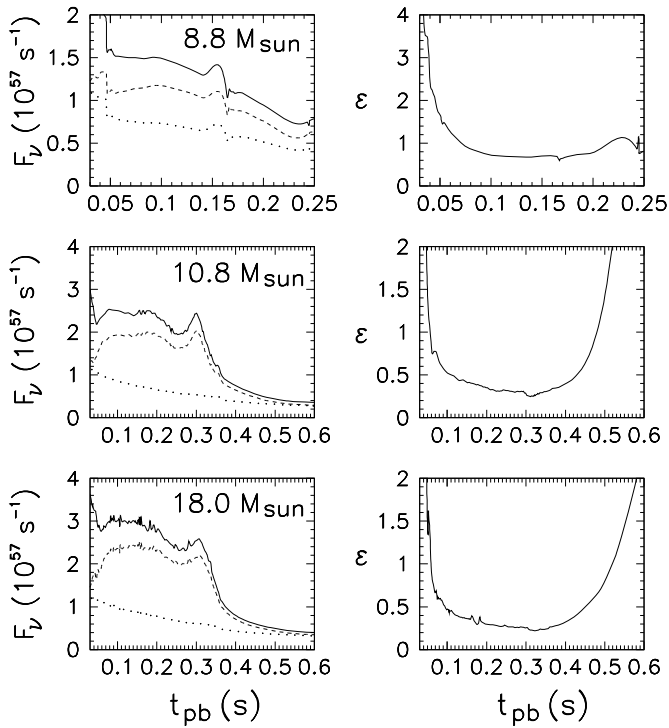}  
\caption{Neutrino particle fluxes (left panels) and 
 flavor asymmetry parameter $\epsilon$ (right panels) for $\nu_e$ (continuous line), ${\overline\nu}_e$ (dashed line),
$\nu_x$ (dotted line).
\label{fig3}}
\end{figure}  

\subsection{Neutrino number fluxes for different flavors}

The neutrino particle flux for the different flavors is defined as
follows 
$F_{\nu_{\alpha}} = {L_{\nu_\alpha}}/{\langle E_{\nu_\alpha} \rangle}$.
In order to characterize the flux asymmetries among the different
$\nu$ species we introduce the parameter~\cite{EstebanPretel:2007ec}
\begin{equation}
\epsilon = \frac{F_{\nu_e}-F_{\overline{\nu}_e}}{F_{\overline{\nu}_e}-F_{\overline{\nu}_x}}
\,\ ,
\label{eq:asym}
\end{equation}
where  we  assume $F_{{\nu}_x}=F_{\overline{\nu}_x}$. 

In Fig.~\ref{fig3} we  show  the neutrino number flux
$F_{\nu_{\alpha}}$ (left panels) and asymmetry parameter $\epsilon$ (right panels)
for the three benchmark SN simulations based on the different progenitor models.
The particle fluxes present the expected hierarchy,
$F_{\nu_e} > F_{{\overline\nu}_e} \gg F_{\nu_x}$,
during the accretion phase.
The first part of the hierarchy is caused by the deleptonization of the collapsed 
stellar core whereas the second is due to the absence of charged-current interactions
for neutrino species other than $\nu_e$ and ${\overline\nu}_e$.
As a consequence of this hierarchy, the electron and non-electron neutrino particle
fluxes differ by almost a factor of  two during the accretion phase.
Conversely, flux differences tend to become much smaller 
during the cooling of the  SN remnant.

Passing to the asymmetry parameter $\epsilon$ [Eq.~(\ref{eq:asym})],
the strong excess of $\nu_e$'s during the early post-bounce deleptonization
($t_{\rm pb} \lesssim 0.05$~s), increases $\epsilon\gg 1$. 
Then, $\epsilon$ drops and reaches values between 0.3-0.5 during the accretion
phase (see Fig.~\ref{fig3}).
After that, since the flux difference $F_{{\overline\nu}_e} - F_{\overline\nu_x}$ drops
more rapidly than $F_{{\nu}_e} - F_{\nu_x}$, $\epsilon$ rises again and becomes
larger than 1 for the iron-core SNe.
We will see that this behavior will have important consequences on the
development of the self-induced flavor conversions.

\subsection{Non-linear neutrino flavor mixing} 

Our description of the non-linear neutrino flavor conversions is based on
the two-flavor oscillation scenario, driven by the atmospheric mass-square difference
$\Delta m^2_{\rm atm} \simeq 2.6 \times 0^{-3}$~eV$^{2}$ and by
a small (matter suppressed) in-medium mixing
$\theta_{\rm eff} = 10^{-3}$~\cite{Fogli:2008ig,Schwetz:2011qt}.
Three-flavor effects, associated with the solar sector,  are  small
for the neutrino flux ordering expected during the accretion
phase~\cite{Dasgupta:2007ws,Mirizzi:2010uz}.

In the normal mass hierarchy and for the spectral ordering of the accretion phase,
no self-induced flavor conversion  will  occur and thus dense
matter  cannot produce any sizable new effect.
Therefore, in the following we will always refer to the inverted mass hierarchy case.

The impact of the non-isotropic nature of neutrino emission on the self-induced flavor
conversions is taken into account by ``multi-angle''
simulations~\cite{Duan:2006an}, where one follows a large number
$[{\mathcal O}(10^3)]$ of neutrino trajectories. 
The $\nu$'s emitted from a SN core naturally have
a broad energy distribution. However, this is largely irrelevant for our purposes, since large matter effects would lock togheter the 
different neutrino energy modes, both in case of  supression~\cite{EstebanPretel:2008ni}  and
of   decoherence~\cite{EstebanPretel:2007ec} of collective oscillations. 
 Therefore, to simplify the complexity of the numerical simulations, we assume all $\nu$'s to
be represented by a single energy, that we take $E=15$~MeV.
This approximation is reasonable since our main purpose is to determine if the
dense matter affects the development of the self-induced transformations.
This choice results in the neutrino vacuum oscillation frequency
\begin{equation}
\omega = \left\langle \frac{\Delta m^2_{\rm atm}}{2E} \right\rangle = 0.4 \,\  \textrm{km}^{-1} \,\ .
\end{equation}

The strength of the neutrino-neutrino interaction, normalized at the neutrinosphere, 
is parametrized by~\cite{EstebanPretel:2007ec}
\begin{eqnarray}
 \mu_r &=& \sqrt{2}G_F \left[n_{{\overline\nu}_e} (r) - n_{{\overline\nu}_x} (r) \right] \nonumber \\
 &=& 7 \times 10^5 \,\ 
\textrm{km}^{-1} \left(\frac{L_{\overline\nu_e}}{\langle E_{\overline\nu_e} \rangle}
-\frac{L_{\overline\nu_x}}{\langle E_{\overline\nu_x} \rangle}
\right) \,\  \nonumber \\
&\times &
\frac{15 \,\ \textrm{MeV}}{10^{52} \textrm{erg}}
\left(\frac{10 \,\ \textrm{km}}{r}\right)^2
\label{eq:mur}
 \end{eqnarray}
 where  $n_{\nu_\alpha}(r)$ is the flux of the neutrino species
 $\nu_\alpha$ at radius $r$
 (for a definition of $n_{\nu_\alpha}$, see Eq.(\ref{eq:nv})).
The matter potential is represented by~\cite{Dighe:1999bi} 
\begin{eqnarray}
\lambda_r &=& \sqrt{2}G_F  n_{e}(r) = 1.9 \times 10^6  \,\ 
\textrm{km}^{-1} \nonumber \\
&\times & \left(\frac{Y_e}{0.5} \right) 
\left(\frac{\rho}{10^{10} \,\ \textrm{g}/\textrm{cm}^{3}} \right) 
\end{eqnarray}
encoding the net $n_e\equiv n_{e^-}- n_{e^+}$  electron density, where
$Y_e=Y_{e^-} - Y_{e^+}$ is the  net  electron fraction and
$\rho$ is the matter density.  
The neutrino term $\mu_r$ declines always as $r^{-2}$ due to the 
geometric dilution  outside the decoupling spheres,
whereas $\lambda_r$ is given by the detailed matter profile from the SN simulations.

\begin{figure}  
\includegraphics[angle=0,width=0.49\textwidth]{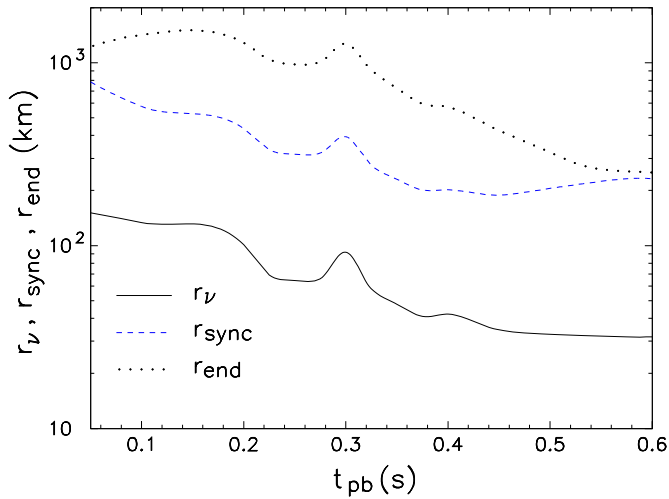}  
\caption{
10.8 M$_{\odot}$ progenitor mass.  Time evolution of $r_{\rm \nu}$ (continuous line),
$r_{\rm sync}$ (dashed curve) and $r_{\rm end}$ (dotted curve).
\label{fig4}}
\end{figure}  

\subsection{Oscillation regimes}

In the  absence of matter suppression, collective neutrino flavor transformations
will  start  outside the synchronization
radius, given by~\cite{EstebanPretel:2007ec}
\begin{equation}
\frac{r_{\rm sync}}{r_\nu} = 
\left(\frac{\sqrt{1+\epsilon} -1}{2}\right)^{1/2} 
\left(\frac{\mu_{r}\big{|}_{r=r_\nu}}{\omega}\right)^{1/4} \,\ , 
\label{eq:rsync}
\end{equation}
and are expected to develop at least until  radii  when the
neutrino-neutrino interaction strength becomes comparable to the vacuum term,
i.e. at~\cite{Fogli:2007bk}
\begin{equation}
\frac{r_{\rm end}}{r_\nu} = \left(\frac{\mu_{r}\big{|}_{r=r_\nu}}{2\omega}\right)^{1/4}  \,\ .
\label{eq:rend}
\end{equation}
We remind that outside  $r_{\rm sync}$ the multi-angle nature
of the neutrino trajectories in a supernova can  lead to
self-induced flavor decoherence between different angular modes.
However, it has been shown in~\cite{EstebanPretel:2007ec} that if the asymmetry
parameter $\epsilon$ is significantly large, multi-angle effects  are
suppressed and self-induced neutrino oscillations exhibit a collective behavior.
We will come back to this point in more details when commenting our numerical results.

Using the above definitions, in Fig.~\ref{fig4} we present  the 
neutrinosphere radius $r_{\nu}$, the synchronization radius $r_{\rm sync}$ and the
radius $r_{\rm end}$ at which collective effects saturate, for the case of the
10.8~M$_{\odot}$ SN simulation.
We see that $r_{\nu} \sim 10^2$~km during the accretion phase and drops to
$\sim 30$~km at the beginning of the cooling phase.
This contraction of the neutrinosphere radius reflects the gradual shift of
the $\nu$ opacities.
These change from being absorption dominated during the accretion, to being
scattering dominated during the cooling~\cite{Janka:1989,Fischer:2011}.

Collective oscillations  are  expected in the range
$r\in[r_{\rm sync},r_{\rm end}]$.
This range is at $r \sim [600,1500]$~km at $t_{\rm pb}=0.1$~s  when
the neutrinosphere radius  $r_\nu >10^2$~km, pushing  to  larger
radii the flavor conversions.
Then, following the contraction of the neutrinosphere radius, the range of conversions
shifts at smaller radii, moving to $r\sim [200,500]$~km at $t_{\rm pb}\simeq 0.4$~s.
Finally at $t_{\rm pb}\simeq 0.6$~s, the lower neutrino luminosity and the larger 
asymmetry parameter $\epsilon$ both conspire to push 
$r_{\rm sync}$ towards $r_{\rm end}$
 [see Eqs.~(\ref{eq:rsync})--(\ref{eq:rend})].

In the range $[r_{\rm sync},r_{\rm end}]$, the self-induced neutrino flavor conversions
can be affected by the dense matter when~\cite{EstebanPretel:2008ni} 
\begin{equation}
 n_{e} \gtrsim n_{\overline{\nu}_e}-n_{\overline{\nu}_x} \,\ .
 \end{equation}
In particular, when the two densities are comparable matter effects induce
multi-angle decoherence among different neutrino modes, leading
to a flavor equilibration among the different species.
When the net electron density is significantly larger than the
neutrino  density,  collective
oscillations are suppressed at all~\cite{EstebanPretel:2008ni}. 
 
We remark that previous numerical studies of self-induced  flavor conversions
typically assumed  smaller  neutrinosphere radii, namely
$r_\nu = {\mathcal O} (10~\textrm{km})$ (see, e.g.,~\cite{EstebanPretel:2007ec}).
This choice is more appropriate for the cooling rather than
for the accretion phase.  
As a consequence of this different input, in the current work flavor conversions start
at larger radii  than  typically  assumed  in previous studies.

\section{Matter effects for different progenitor masses} 

In this Section we will present our results concerning the effects of dense matter on
of the self-induced neutrino oscillations for the three representative SN simulations
based on the different progenitor masses.

\subsection{10.8 M$_\odot$ progenitor mass}

\begin{figure*}  
\includegraphics[angle=0,width=0.8\textwidth]{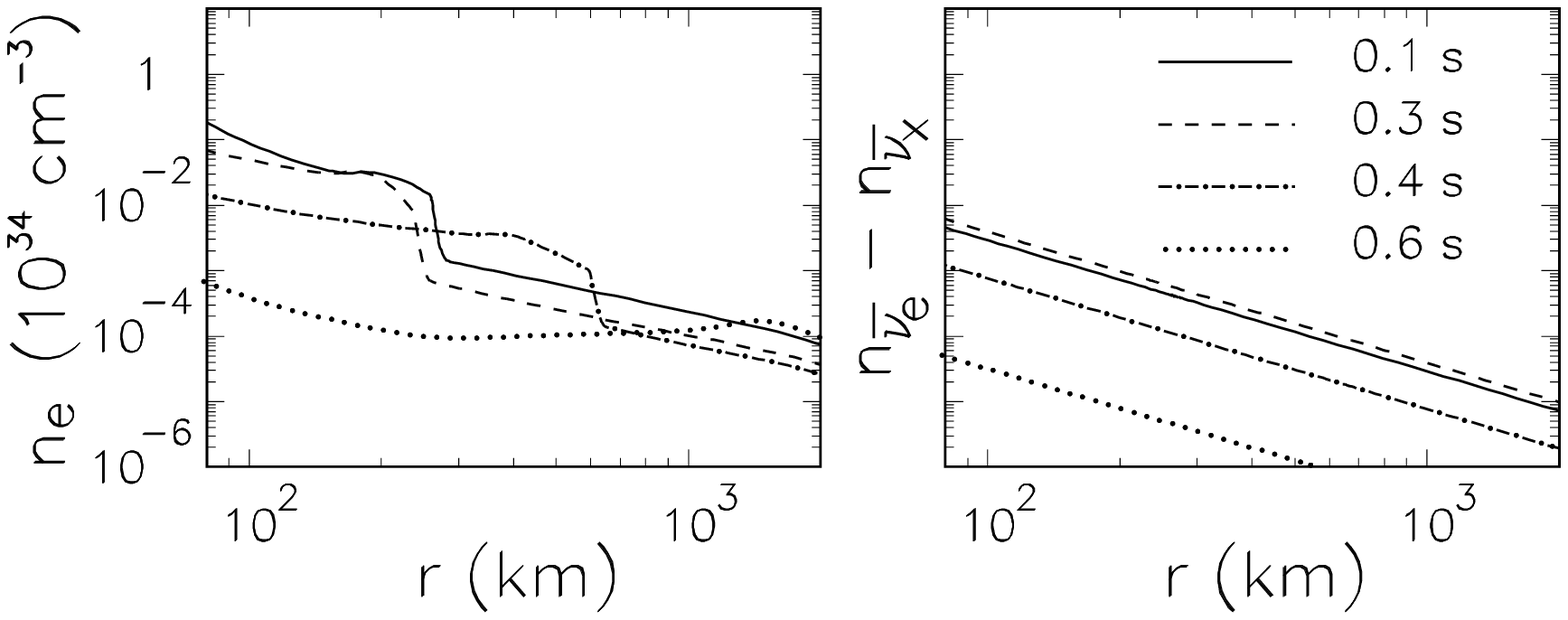}  
\caption{
10.8 M$_{\odot}$ progenitor mass.  Radial evolution of the net electron density
$n_e$ (left panel) and of the neutrino density difference
$n_{{\overline\nu}_e}- n_{{\overline\nu}_x}$ (right panel)
at different post-bounce times.
\label{fig5}}
\vskip24pt
\includegraphics[angle=0,width=0.8\textwidth]{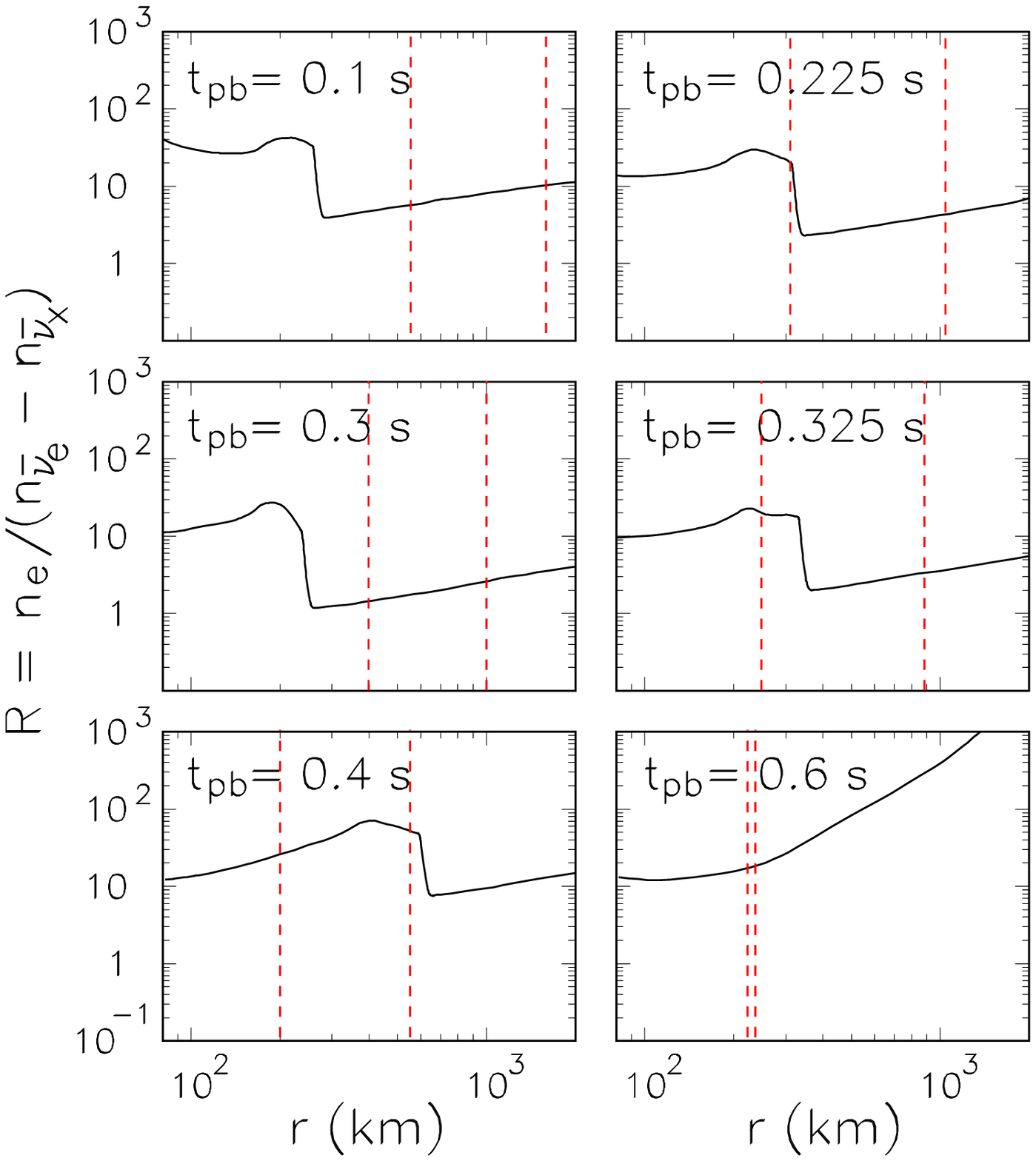}  
\caption{
10.8  M$_{\odot}$ progenitor mass. Radial evolution of the ratio $R$ between
electron and neutrino densities at different post-bounce times.
The two dashed vertical strips delimit the position of $r_{\rm sync}$ (left line)
and $r_{\rm end}$ (right line).}
\label{fig6}
\end{figure*}  

We start our investigation with the case of  the 10.8~M$_\odot$ iron-core SN. 
In Fig.~\ref{fig5} we  show  the net electron density $n_e$
(left panel) and the difference of neutrino densities
$n_{\overline\nu_e}-n_{\overline\nu_x}$ (right panel) entering in the potential $\mu_r$ 
[Eq.~(\ref{eq:mur})] for different post-bounce times. 
While the decline of the neutrino density is $r^{-2}$, the electron density presents
a more complicated behavior.
In particular, one can recognize the abrupt discontinuity in $n_e$ associated
 with  the supernova shock-front that propagates in time.
The shock-wave initially dissipates its energy due to heavy nuclei dissociation
and eventually stalls at $t_{\rm pb}=0.05$~s.
The standing accretion shock is revived via neutrino heating, on a timescale
on the order of 0.1~seconds, and expands accordingly as well as contracts via
neutrino cooling.
After 0.35~seconds post-bounce, neutrino heating succeeds and the standing
accretion shock turns into a dynamic shock with positive matter velocities.
It initiates the onset of explosion at about 0.4~seconds post-bounce.

For this  SN model  we find that during the accretion phase 
the matter density in the post-shock region declines slower than the neutrino density, 
typically as $\sim r^{-1.5}$.
From the comparison between the electron and the
neutrino densities, we realize that at the different post-bounce
times considered, $n_e$ is always larger than or comparable to
$n_{\overline\nu_e}-n_{\overline\nu_x}$.
It suggests that one cannot ignore matter effects on self-induced flavor
transformations during the accretion phase. 
 
In order to quantify the relative strength of the electron and neutrino densities,
in Fig.~\ref{fig6} we show the ratio
\begin{equation}
R= \frac{n_{e}}{n_{\overline{\nu}_e}-n_{\overline{\nu}_x}} 
 \,\ .
\label{eq:ratio}
\end{equation}
 as a  function of  the radial coordinate
$r$ at different post-bounce times  for $t_{\rm pb} \in [0.1,0.6]$~s.
The range $r_{\rm sync} < r < r_{\rm end}$
is delimited with two vertical dashed lines. 
The values of $r_{\rm sync}$ and $r_{\rm end}$ determine the possible range for the
self-induced flavor conversions and the shock radius $r_{\rm sh}$ denotes the abrupt
drop in the electron density.
Therefore, their relative position is crucial to assess the impact of matter effects.
In the expected oscillation range,   $R\gg 1$  will  imply a strong matter
dominance in the flavor conversions and thus complete suppression of the
self-induced effects.
Instead, when electron and neutrino densities  are  comparable
($R\gtrsim 1$), decoherence  will  occur for the collective
oscillations.

The ratio $R$, being very large behind the shock front, prevents flavor conversions
in this region. 
However, the ratio can go down to $R\gtrsim 1$ for $r>r_{\rm sh}$, 
leading to matter-induced decoherence and thus partial flavor changes.

Let us discuss in more detail what occurs at different post-bounce time snapshots
in Fig.~\ref{fig6}.
At very early times ($t_{\rm pb}=0.1$~s) the matter term is strongly
dominant also behind  the shock-front ($R\gg 1$).
Under these conditions oscillations are always blocked.
Then, at intermediate times ($t_{\rm pb}=0.225, 0.3$~s)  the matter density in the
post-shock region, where flavor conversions  are  possible,
is dropping faster than the neutrino one.
Therefore, the ratio $R$ drops at 1-2  in this range and
 matter-induced decoherence is possible in this case.
Subsequently, at $t_{\rm pb}=0.325$~s, oscillations  are
suppressed  behind  the shock front, but then decoherence
 will  develop at larger radii ($r\gtrsim 300$~km) when
$R \gtrsim 1$.  
Eventually at later times ($t_{\rm pb}=0.4, 0.6$~s), since the shock has resumed
its forward motion, the region relevant for the oscillations  is  at 
$r<r_{\rm sh}$, where $R\gg 1$.
In this situation self-induced oscillations will be suppressed.
From these different snapshots we realize that $R$ has a peculiar non-monotonic
behavior  as a  function of time.
 It  suggests a time-dependent pattern for the matter effects
on the self-induced transitions during the accretion phase,
namely complete--partial--complete suppression.
 
\begin{figure}  
\includegraphics[angle=0,width=0.49\textwidth]{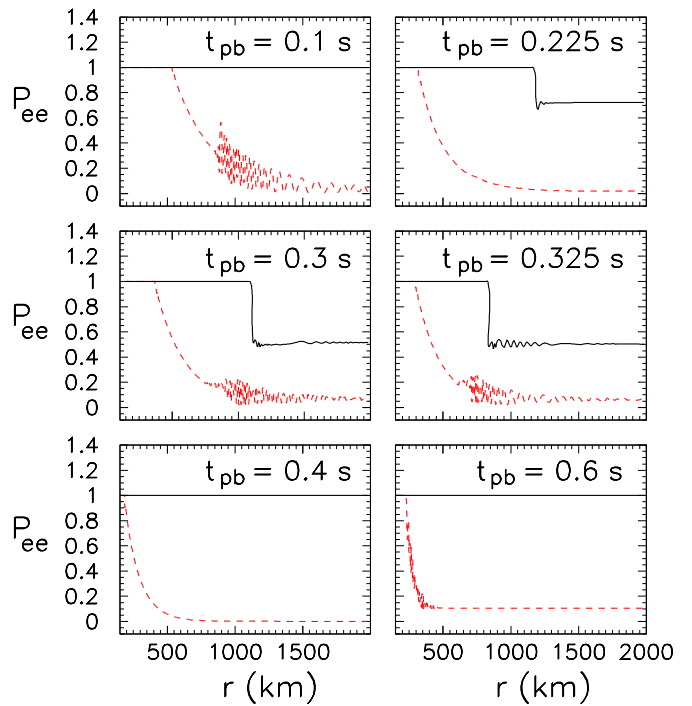}  
\caption{
10.8 M$_{\odot}$ progenitor mass. Radial evolution of the survival probability $P_{ee}$
for electron antineutrinos  at different post-bounce times for the multi-angle
evolution 
in presence of matter effects (continuous curve) and for $n_e=0$ (dashed curve).}
\label{fig7}
\end{figure}  

\begin{figure}  
\includegraphics[angle=0,width=0.4\textwidth]{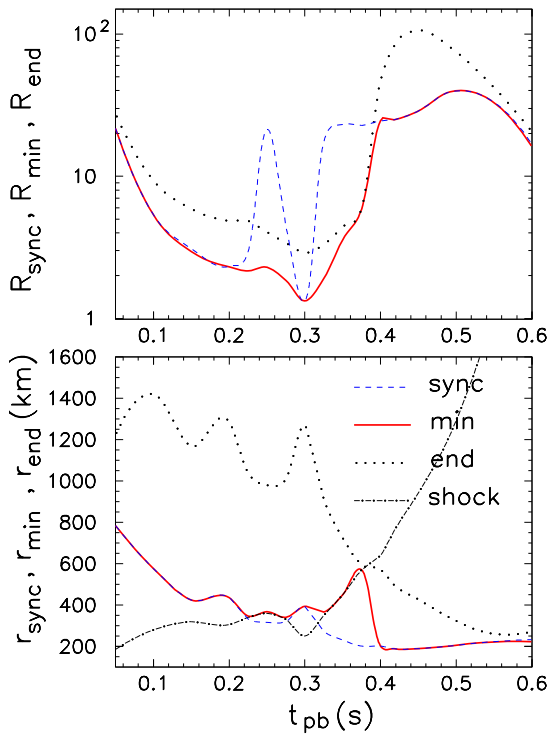}  
\caption{
10.8 M$_{\odot}$ progenitor mass. 
Upper panel: Time evolution of the ratio $R_{\rm sync}$ (dashed curve),
$R_{\rm min}$ (continuous curve) and $R_{\rm end}$ (dotted curve). 
Lower panel: Time evolution of the radial position $r_{\rm sync}$ (dashed curve),
$r_{\rm min}$ (continuous curve), $r_{\rm end}$ (dotted curve) and
 $r_{\rm shock}$ (dot-dashed curve). }
\label{fig8}
\end{figure}  

In order to confirm these expectations, we have performed a multi-angle numerical
study of the equations for the neutrino flavor evolution in the schematic model 
described in Section~III.
Our treatment closely follows the one presented  in Ref.~\cite{EstebanPretel:2008ni}
to which we address the interested reader for further details. 
We only mention here that in order to achieve convergence in our simulations we
had to simulate $10^3$ neutrino angular modes. 
In Fig.~\ref{fig7} we show the radial evolution of the ${\overline\nu}_e$ survival
probability $P_{ee}$ for  different post-bounce times, obtained taking into account the
effects of the SN matter profile (continuous curve).
For comparison, we show also the results  obtained setting $n_e=0$ (dashed curve).
 
In the case with $n_e=0$, for the given flavor asymmetry $\epsilon \gtrsim 0.3$ we would have expected the ``quasi-single angle'' behavior described
in Ref.~\cite{EstebanPretel:2007ec}, where after the onset of the conversions at
$r=r_{\rm sync}$, the survival probability $P_{ee}$ declines smoothly approaching
zero at large radii.
However, in the situation we are studying flavor conversions develop at  radii
larger than what  is  typically shown in previous works
(see, e.g.,~\cite{EstebanPretel:2007ec}). Therefore, the evolution is more adiabatic
(i.e. the evolution length scale $l_{\mu} \sim r$~\cite{Dasgupta:2010cd}).
As a consequence, effects of self-induced multi-angle decoherence have more 
chances to develop in this case, producing some small disturbance in the smooth
decline of the survival probability  at large radii
(visible at $t_{\rm pb}=0.1, 0.3, 0.325$~s for $r\gtrsim 700$~km).
This finding is potentially interesting, however, since matter effects will anyhow
dramatically alter this picture we have not  performed a systematic study on this
(sub-leading) self-induced decoherence. 

Passing now to the matter case, we see that at $t_{\rm pb}=0.1, 0.4, 0.6$~s the flavor
oscillations are completely blocked, since $R\gg 1$ in the conversion range. 
For the other three intermediate times ($t_{\rm pb}=0.225, 0.3, 0.325$~s), the presence
of a large matter term at $r_{\rm sync}$ will significantly delay the onset of the flavor
conversions with respect to the case with $n_e=0$.
Then, at larger radii ($r >700$~km) when $R\gtrsim$1--2, matter effects produce a
partial suppression of the flavor conversions, with a tendency toward  flavor decoherence.
When the latter is complete, it leads to $P_{ee} \to 1/2$ (at $t_{\rm pb}=0.3, 0.325$~s),
implying a complete mixture between ${\overline\nu}_e$ and ${\overline\nu}_x$.

Finally, we would  summarize our results  with a figure of merit that captures in a compact
way the interplay between matter and self-induced effects as a function of time. Therefore, 
in Fig.~\ref{fig8} we show the ratio $R$ (upper panel) at its minimum value
\begin{equation}
R_{\rm min} = \textrm{min}\left(\frac{n_{e}}{n_{\overline{\nu}_e}-n_{\overline{\nu}_x}} 
\right)_{r \in [r_{\rm sync}; r_{\rm end}]} \,\ ,
\label{eq:ratio}
\end{equation}
in the range $r \in [r_{\rm sync}; r_{\rm end}]$ relevant for collective conversions. 
This value provides a conservative estimation for the impact of matter effects. 
Indeed, if $R_{\rm min} \gg 1$ one should expect a complete matter suppression of 
the flavor changes.
Conversely, when $R_{\rm min} \gtrsim 1$ the matter suppression will be partial
and will lead to flavor decoherence. 
For comparison we also show the values of $R$ at the two radii $r_{\rm sync}$ and
$r_{\rm end}$.
Moreover, in the lower panel we represent the time evolution of the radial positions
$r_{\rm sync}$, $r_{\rm end}$, $r_{\rm min}$ and $r_{\rm sh}$.

One realizes that  $R$ presents a non-trivial behavior in the range between
$r_{\rm sync}$ and $r_{\rm end}$. 
In particular, the position   of the  minimum value $R_{\rm min}$ is at
\begin{equation}
r_{\rm min} = \left\{\begin{array}{rl}
r_{\rm sh} & \mbox{if} \,\ \,\  r_{\rm sync}< r_{\rm sh}< r_{\rm end} \\
r_{\rm sync}   & \mbox{otherwise}
\end{array}
\right .
\end{equation}
For $t_{\rm pb}\lesssim 0.3$~s, since  the neutrino density grows and the electron
density decreases in time, $R_{\rm min}$ decreases monotonically from $\sim$20
at $t_{\rm pb}=0.05$~s to $\sim$1--2 at $t_{\rm pb}\simeq 0.3$~s post-bounce,
changing  from complete to partial matter suppression.
Then, since at the end of the accretion  phase  the neutrino
density decreases faster than the electron  density,
$R_{\rm min}$ rises again to $\sim$20 at $t_{\rm pb}\simeq 0.4$--0.6~s,
suppressing flavor conversions. 
The behavior of $R_{\rm min}$ shows in a compact way the succession of phases with
complete--partial--complete matter suppression in the time evolution of the neutrino
self-induced conversions during the  accretion phase.
   
A similar analysis to  the one
presented  above  based on Fig.~\ref{fig8}, 
 may  allow to obtain a first answer about the role
of the matter on the self-induced effects.
This schematic approach does not require a detailed numerical study of the
neutrino flavor evolution.
For this reason, we find it useful to compare the role of matter in different SN
simulations based on different progenitor models.
In the following, we will apply such analysis to the other SN simulations from
Ref.~\cite{Fischer:2009af} based on different progenitor masses.
 
\begin{figure*}  
\includegraphics[angle=0,width=0.8\textwidth]{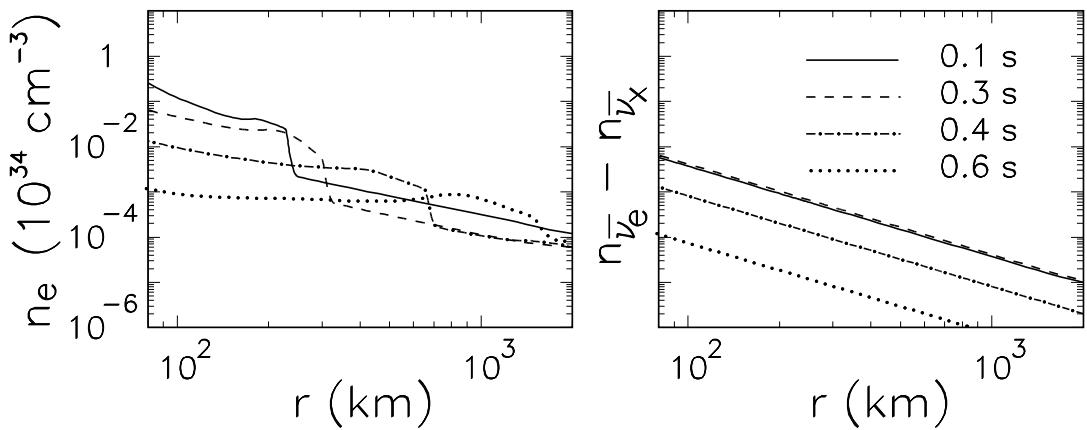}  
\caption{
18.0 M$_{\odot}$ progenitor mass.  Radial evolution of
the net electron density $n_e$ (left panel) and of the neutrino
density difference $n_{\overline\nu_e}- n_{\overline\nu_x}$ (right panel) at different
post-bounce times.
\label{fig9}}
\end{figure*}  

\begin{figure}
\includegraphics[angle=0,width=0.4\textwidth]{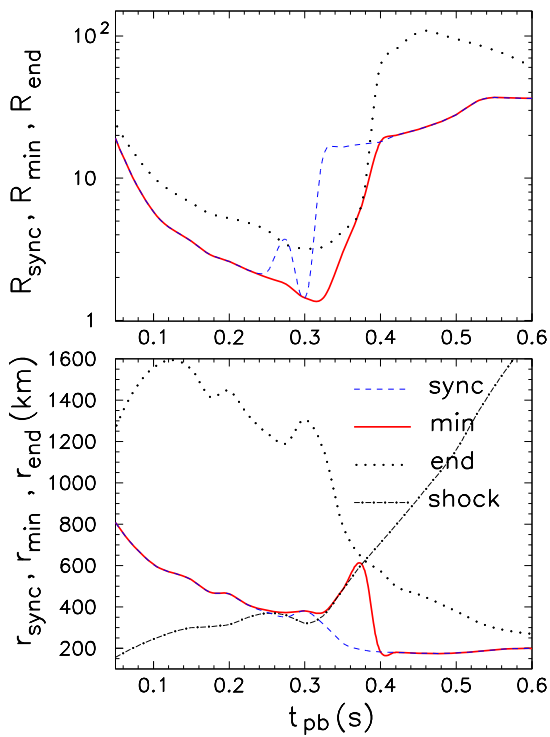}  
\caption{
18.0 M$_{\odot}$ progenitor mass.  Upper panel: Time evolution of the ratio $R_{\rm sync}$ (dashed curve),
$R_{\rm min}$ (continuous curve) and $R_{\rm end}$ (dotted curve). 
Lower panel: Time evolution of the radial position $r_{\rm sync}$ (dashed curve),
$r_{\rm min}$ (continuous curve), $r_{\rm end}$ (dotted curve) and
 $r_{\rm shock}$ (dash-dotted curve). 
\label{fig10}}
\end{figure}

\subsection{18.0 M$_\odot$ progenitor mass}

As further example, we consider the SN simulation of the
more massive  18~M$_\odot$ iron-core progenitor.
In Fig.~\ref{fig9} we show the net electron density $n_e$ (left panel) and the
difference of neutrino densities $n_{\overline\nu_e}-n_{\overline\nu_x}$  for
different post-bounce times. 
From the comparison with Fig.~\ref{fig5} we see that the time evolution of the
electron and neutrino density profiles  is  similar  to
the case 10.8~M$_\odot$ SN simulation.
Therefore, we expect a similar impact for the matter effects  on the neutrino
flavor evolution. 
In Fig.~\ref{fig10} we show the time evolution of $R_{\rm min}$ in the same format
 as in  Fig.~\ref{fig8}.
We see that the behavior of $R_{\rm min}$ is  also very similar
 to the previous case described above.
Therefore, one expect an analogous pattern of complete 
$0.05\lesssim t_{\rm pb} \lesssim 0.2$~s),
partial ($0.2\lesssim t_{\rm pb} \lesssim 0.35$~s)
and complete ($0.35\lesssim t_{\rm pb} \lesssim 0.6$~s) matter suppression in the self-induced
flavor conversions.

We have repeated the same analysis also for a 15.0 M$_\odot$ iron-core progenitor mass,
giving a not-exploding supernova. In the time range before the recollapse of the star
($t_{\rm pb} \lesssim 0.3$~s) the results obtained are similar to the one shown for the two exploding
cases. Therefore, for the sake of the brevity, we will not show them here.

\subsection{8.8 M$_\odot$ progenitor mass}

\begin{figure*}  
\includegraphics[angle=0,width=0.8\textwidth]{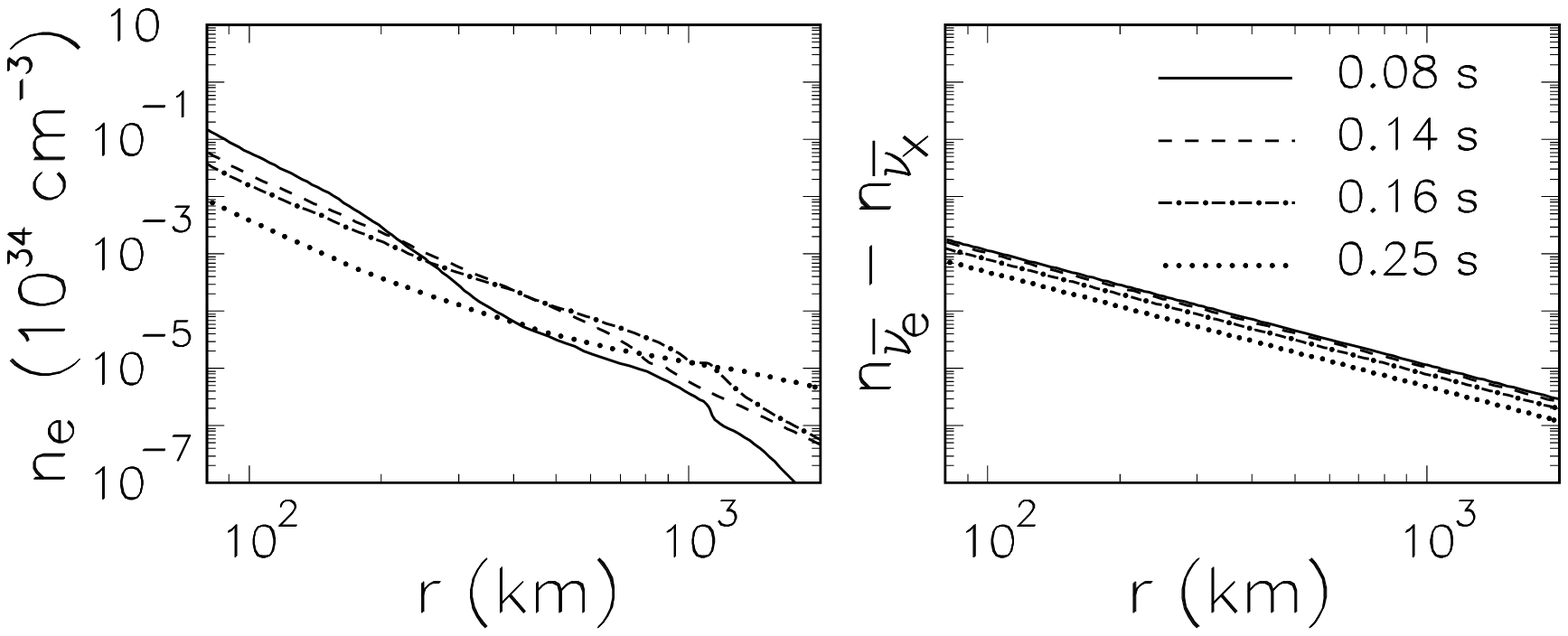}  
\caption{
8.8 M$_{\odot}$ progenitor mass.  Radial evolution of
the net electron density $n_e$ (left panel) and of the neutrino
density difference $n_{\overline\nu_e}- n_{\overline\nu_x}$ (right panel) at different
post-bounce times. 
\label{fig11}}
\end{figure*}  

\begin{figure} 
\includegraphics[angle=0,width=0.4\textwidth]{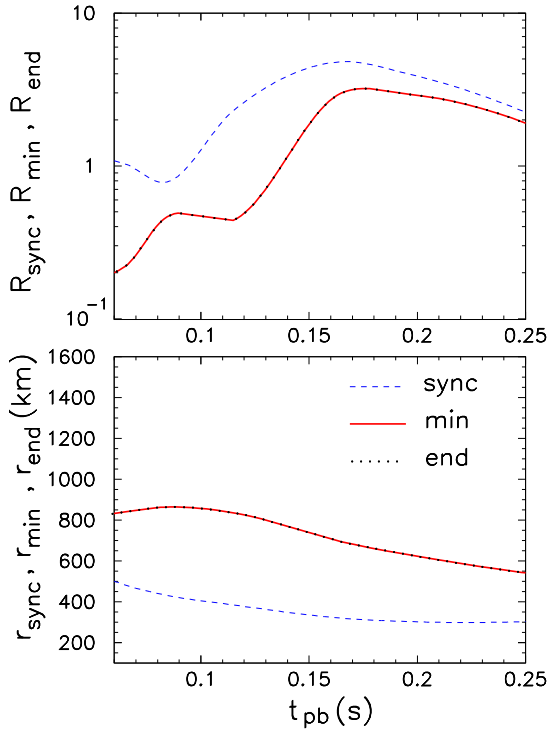}  
\caption{
8.8 M$_{\odot}$ progenitor mass. 
Upper panel: Time evolution of the ratio $R_{\rm sync}$ (dashed curve),
$R_{\rm min}$ (continuous curve) and $R_{\rm end}$ (dotted curve). 
Lower panel: Time evolution of the radial position $r_{\rm sync}$ (dashed curve),
$r_{\rm min}$ (continuous curve), $r_{\rm end}$ (dotted curve).\label{fig12}}
\end{figure}

Finally, we pass to analyze the case of the SN of the low mass 8.8~M$_\odot$
O-Ne-Mg core.
In Fig.~\ref{fig11} we present the net electron density $n_e$ (left panel) and the
difference of neutrino densities $n_{\overline\nu_e}-n_{\overline\nu_x}$ (right panel)
for different post-bounce times. 
We realize that since in this case the matter density of the 
envelope is very low compared to the iron-core progenitors,
the electron density profile above the core is very steep,
declining as $\sim r^{-2.5}$, faster than the neutrino density. 
Moreover, also the neutrino densities for $t_{\rm pb}\lesssim 0.2$~s are
roughly a factor $\sim 3$ smaller than in the case of the iron-core supernovae.
This reflects the practical absence of an extended accretion
phase for this low-mass star. 
In this case the explosion succeeds very shortly after the core-bounce.
Therefore,  the shock-front  is  already beyond the radial range
interesting for the flavor conversions.

\begin{figure}  
\includegraphics[angle=0,width=0.49\textwidth]{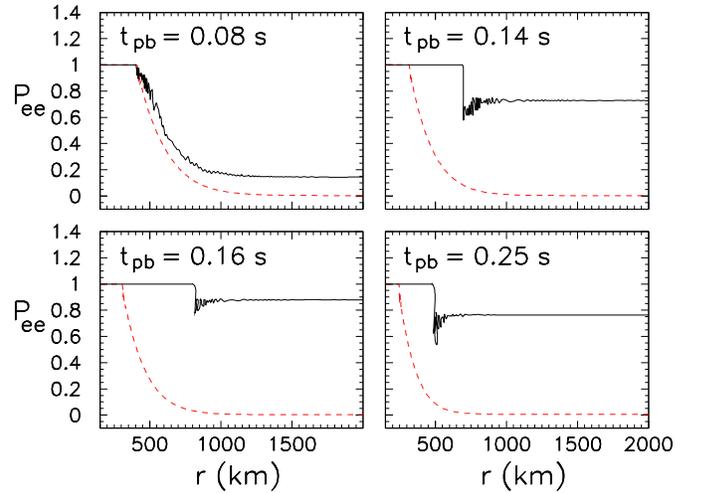}  
\caption{
8.8  M$_{\odot}$ progenitor mass. Radial evolution of the survival probability 
$P_{ee}$ for electron antineutrinos at different post-bounce times for the multi-angle
simulations in presence of matter effects (continuous curve)  and for the case with
 $n_e=0$ (dash-dotted curve). 
\label{fig13}}
\end{figure}  

Fig.~\ref{fig12} shows the time evolution of $R_{\rm min}$ in the range
$r\in[r_{\rm sync}; r_{\rm end}]$ relevant for collective conversions.
We realize that the  behavior of $R_{\rm min}$ is different with respect 
to the previously considered iron-core supernovae. 
Since in this case the matter does not present an abrupt discontinuity, but
it monotonically decreases, the position of $R_{\rm min}$ will always coincide with
$r_{\rm end}$.
Moreover, since the electron density is never much larger than the neutrino density
in the conversion region, it results  in  $R_{\rm min} \lesssim 3$.

The peculiar time evolution of $R_{\rm min}$ in this case will depend on 
the small differences between the matter and the neutrino term.
At $t_{\rm pb}\lesssim 0.13$~s, the electron density is smaller  than the neutrino
density in the range of collective conversions, giving $R_{\rm min} \lesssim 1$.
Then, the ratio $R_{\rm min}$ grows  up to  $\sim 2-3$ at 
$t_{\rm pb}\simeq 0.2$~s since the  electron density in the
conversion region is larger than in the previous case. 
As a result, the  time-evolution of the neutrino survival probabilities presents a
change between a regime dominated by the $\nu$--$\nu$ effects at very early times
and one of matter-suppressed oscillations at later times.   

In Fig.~\ref{fig13} we explicitly show this behavior, representing the radial evolution
of the  ${\overline\nu}_e$  survival probability $P_{ee}$ for the same post-bounce
times snapshots of Fig.~\ref{fig11}, with (continuous curves) and without (dashed curves)
matter. 
We  find  that at $t_{\rm pb}=0.08$~s where $R\lesssim 1$,
the matter suppression is relatively small and the neutrino-neutrino interactions
produce a quite almost complete swap between  ${\overline\nu}_e$ and
${\overline\nu}_x$ spectra ($P_{ee}=0.15$ at the end of the evolution).
Conversely, for the  later  times considered, the flavor conversions
are strongly suppressed with a final $P_{ee}\simeq 0.7-0.9$, since $R\gtrsim 1$.
We want to remark that the time evolution of the neutrino oscillation probability during the 
accretion phase is significantly different with respect to the case of the iron-core
supernovae we studied.
This will allow the distinction of a O-Ne-Mg-core SN from a
iron-core SN, in the case of the detection of a future galactic event.
%

\section{Earth matter effect during the accretion phase}

It is plausible that a high-statistics  neutrino detection from a future
galactic SN~\cite{Autiero:2007zj,Wurm:2011zn} would faithfully monitor the abrupt
time variations imprinted on the neutrino oscillation probabilities by the transitions
between stages of complete and partial matter suppression  during the accretion phase.
We plan to perform a dedicated investigation of the matter suppression during the
accretion phase on the observable SN neutrino burst in a future work.

Here we discuss another consequence of the SN dense matter effects, namely 
the change  of  the interpretation of the Earth matter effect
on the supernova neutrino signal.
It is widely known that if SN neutrinos arrive at the detector from ``below'', the Earth
crossing  will  induce an energy-dependent modulation in the
neutrino survival probability~\cite{Lunardini:2001pb}.
The appearance of the Earth effect depends on the neutrino fluxes and on the
mixing scenario.
The accretion phase is particularly promising to detect Earth crossing signatures
because the absolute SN $\nu$ flux is large and the flavor-dependent 
flux differences are also large.

Before the inclusion of the collective effects in the SN neutrino flavor transitions, 
the detection of Earth matter modulations has been proposed as a tool to distinguish
the neutrino mass hierarchy at ``large'' values of the mixing angle
$\theta_{13}$ (i.e. $\sin^2\theta_{13} \gtrsim 10^{-3}$)~\cite{Dighe:2003jg}. 
Then, when collective oscillations were taken into account, it was speculated
that due to the self-induced flavor transformations, the presence or absence
of Earth matter effects  will break the degeneracy between
normal and inverted mass hierarchy for ``small'' values of the mixing angle
$\theta_{13}$ (i.e. $\sin^2\theta_{13} \lesssim 10^{-5}$) while no hierarchy
discrimination  is  possible for larger values of
$\theta_{13}$~\cite{Dasgupta:2008my}.
Now, due to the matter suppression effect this conclusion should be revisited
for the accretion phase. 

To determine the observable SN neutrino fluxes at Earth, we
refer to~\cite{Dasgupta:2007ws}.
We consider the modified flavor basis,
$(\nu_e, \nu_x, \nu_y)$,
which is  defined such that
$(\nu_e, \nu_x, \nu_y) = {R_{23}}^\dagger (\theta_{23})  
(\nu_e, \nu_\mu, \nu_\tau)$,
where $R_{23}$ is the 2--3 rotation matrix.
This will allow for  collective oscillations between $e$
and $y$ states.
For definiteness, here we concentrate on the ${\overline\nu}_e$ spectra observable
through inverse beta decay reactions,
${\overline\nu}_e + p \longrightarrow n + e^+$,
at water/ice Cherenkov or scintillation detectors~\cite{Autiero:2007zj,Dighe:2003be}.

In normal hierarchy, the ${\overline\nu}_e$ flux at Earth, $F_{{\overline\nu}_e}^D$, 
(without collective effect and no MSW effect associated to $\theta_{13}$)
for any value of the mixing angle $\theta_{13}$  is  given
by~\cite{Dighe:2003jg}
\begin{equation}
F_{{\bar\nu}_e}^D = \cos^2 \theta_{12} F_{{\bar\nu}_e} + 
\sin^2 \theta_{12} F_{{\bar\nu}_x} \,\ , 
\label{eq:nhearth}
\end{equation}
where $\theta_{12}$ is the 1--2 mixing angle, with 
$\sin^2 \theta_{12}  \simeq 0.3$~\cite{Fogli:2008ig,Schwetz:2011qt}. 

In the inverted mass hierarchy, MSW matter effects in the SN
envelope are characterized in terms of the level-crossing probability $P_H$
of antineutrinos, which is in general a function of the neutrino energy and
$\theta_{13}$~\cite{Dighe:1999bi}.
In the following, we consider two extreme limits, namely $P_H \to 0$ when
$\sin^2 \theta_{13} \gtrsim 10^{-3}$ (``large''), and $P_H \to 1$ when
$\sin^2\theta_{13} \lesssim 10^{-5}$ (``small'').
In this situation  and for complete matter suppression of collective
oscillations for large $\theta_{13}$ , $F_{{\bar\nu}_e}^D$ is given by
\begin{equation}
F_{{\bar\nu}_e}^D =F_{{\bar\nu}_x} \,\ ,
\label{eq:ih}
\end{equation}
while  small $\theta_{13}$ results in the same case as for normal hierarchy [Eq.~(\ref{eq:nhearth})].

In the case of  matter-induced  docoherence between ${\overline\nu}_e$ 
and ${\overline\nu}_y$ in inverted hierarchy, leading to a complete   mixture of 
${\overline\nu}_e$ and ${\overline\nu}_y$,  the detectable ${\overline\nu}_e$ flux for any value of the mixing angle
 $\theta_{13}$ would be 
given by
\begin{equation}
F_{{\bar\nu}_e}^D = \cos^2 \theta_{12} \frac{F_{{\bar\nu}_e}+F_{{\bar\nu}_y}}{2} + 
\sin^2 \theta_{12} F_{{\bar\nu}_x} \,\ . 
\label{eq:ihearthsmall}
\end{equation}

Earth effect can be taken into account by just mapping
$\cos^2 \theta_{12} \to P({\overline\nu}_1 \to {\overline\nu}_e)$ and
$\sin^2 \theta_{12} \to 1-P({\overline\nu}_1 \to {\overline\nu}_e)$.
where $P({\overline\nu}_1 \to {\overline\nu}_e)$ is the probability that a state entering
the Earth as mass eigenstate ${\overline\nu}_1$ is detected as ${\overline\nu}_e$ at the
detector~\cite{Lunardini:2001pb}.

Considering  for definiteness the case of iron-core SNe, the presence or absence
of Earth matter effects at early times ($t_{\rm pb}\lesssim 0.2$~s)
 will  allow to distinguish the neutrino mass hierarchy at large
value of the mixing angle $\theta_{13}$ [see Eqs.~(\ref{eq:nhearth})--(\ref{eq:ih})]. 
Moreover,  the Earth effects may become observable via the interesting
time pattern introduced above, in the case of inverted mass hierarchy at
large $\theta_{13}$:
\emph{(1)} no effects at early times  [Eq.~(\ref{eq:ih})],
\emph{(2)} the appearance at intermediate times when decoherence occurs
[Eq.~(\ref{eq:ihearthsmall})], and
\emph{(3)} disappearance again when the accretion rate decreases significantly.

\section{Discussion and Conclusions}

Different simulations of core-collapse SNe ~\cite{shock,Tomas:2004gr,Buras:2005rp, Liebendoerfer:2003es},
show  independently  that the matter density in the stellar interior is large during the
post-bounce accretion phase  before the onset of an explosion.
This implies that self-induced neutrino flavor transformations are affected
by the high matter density, through trajectory-dependent
phenomena~\cite{EstebanPretel:2008ni}.

In order to characterize the SN $\nu$ flavor evolution quantitatively, we performed a dedicated
study, where we took as benchmark for the SN neutrino emissivity and the matter
profiles the recent long-term core-collapse SN simulations from
Ref.~\cite{Fischer:2009af}. 
We  considered three different cases for the supernova progenitor
mass, the low mass 8.8~M$_\odot$ O-Ne-Mg-core and the two more massive
iron-core progenitors of 10.8 and 18~M$_\odot$.
In all three cases the electron density $n_e$ is never negligible in comparison
to the neutrino density $n_{\nu}$ during the accretion phase.
As a consequence, the trajectory-dependent matter effects always influence
the development of the self-induced transformations. 
We realized that during the accretion phase both the condition of matter suppression
($n_{e} \gg n_{\nu}$) and matter-induced decoherence ($n_{\nu} \sim  n_e$) 
 are  realized.
This implies  that interesting time-dependent features
in the neutrino signal can occur  at early times.
Moreover, the different patterns of complete/partial matter suppression would allow, 
at least in principle, to distinguish  from the flavor evolution between iron-core
and O-Ne-Mg core SNe.

In early schematic investigations, 
the accretion phase seemed a particularly clear setup to probe the
development of the collective neutrino transformations, because the expected flux
hierarchy is large and robust.
In such a situation the pattern of the self-induced spectral swaps/splits were thought
to be unambiguous~\cite{Mirizzi:2010uz}.
In this scenario, the presence or absence of Earth matter effects on the SN neutrino
burst were proposed to break the degeneracy between normal and inverted mass
hierarchy for ``small'' values of the mixing angle $\theta_{13}$~\cite{Dasgupta:2008my}.
Our findings seem to change this picture once more.
In particular, the dense matter suppression of collective oscillations in supernovae
 implies that Earth matter effects  may 
allow to distinguish the neutrino mass hierarchy in the likely case 
of large $\theta_{13}$~\cite{Abe:2011sj},
as originally  expected in the analysis performed
 without  the inclusion of the collective 
phenomena.
This intriguing possibility  makes  supernova neutrino detection
a competitive way to get this  yet \ unknown property of the
neutrino mass spectrum in addition to terrestrial  experiments~\cite{Mezzetto:2010zi}.

Furthermore, in the past~\cite{shock,Duan:2006an} it has been speculated that swaps
of neutrino fluxes between electron and non-electron flavors in the deepest SN regions
may increase the neutrino energy deposition and hence the neutrino
heating in the gain region behind the standing bounce shock.
It was argued that it could possibly help to revive the shock and to trigger
neutrino-driven explosions.
We have shown that the presence of the matter suppression of flavor
conversions at high densities, behind the shock front
(see, e.g., Fig.~6), implies that these cannot play any significant role in
changing the neutrino energy deposition rate behind the stalled shock-wave.
As an important consequence, the problem of he $\nu$ flavor mixing in SNe can be
decoupled from the $\nu$ transport and impact on the  matter heating/cooling.
This result is confirmed by the analysis recently performed in~\cite{Dasgupta:2011jf}
using two-dimensional supernova models. In that work, even if multi-angle matter effects are not included
in the simulation of the flavor oscillations, a negligible impact of the self-induced conversions is 
found on the energy deposition behind the shock-wave. Indeed, we also find that neglecting matter
effects, the possible range where self-induced oscillations would have significantly developed is
always after the shock-wave position ($r_{\rm sh} \lesssim r_{\rm sync}$) at the times relevant for the
shock revival (see our Fig.~6--8).

Our results have been obtained considering a spherically symmetric neutrino emission. 
All the previous analysis in the field have relied on this assumption  to make
the flavor evolution equations numerically tractable.  It remains to be investigated if the 
removal of a perfect spherical symmetry can provide a different behavior in the flavor 
evolution~\cite{Sawyer:2008zs}.
Moreover, in multi-dimensional SN models density fluctuations are expected behind the standing 
bounce shock, due to the presence of convection and hydro instabilities. These can range  at 
most between 10$\%$ to a factor 2-3 (see, e.g.,~\cite{Tomas:2004gr,Scheck:2006rw}). Therefore,
even in the case a fraction of neutrinos beam crosses underdenses regions, the matter suppression of the collective oscillations will still remain relevant. 
This claim is supported by a recent analysis of the matter suppression,  performed with two-dimensional
SN simulations~\cite{Dasgupta:2011jf}.

Self-induced effects in the supernova neutrinos still remain crucial during the later
cooling phase,  when the matter density decreases continuously 
($t_{\rm pb}\gtrsim 1$~s) and becoming sub-dominant with respect to the neutrino density.
In this situation and for the flux ordering provided by the  simulations from Ref.~\cite{Fischer:2009af}, we confirm
that self-induced oscillations in the cooling phase produce multiple
splits and swaps,
like the one described
in~\cite{Dasgupta:2009mg,
Fogli:2009rd,
Chakraborty:2009ej,
Friedland:2010sc,
Dasgupta:2010cd,
Duan:2010bf,
Mirizzi:2010uz}.
Moreover, for low-mass O-Ne-Mg-core SN a peculiar interplay of MSW and
self-induced effects would produce interesting signatures during the $\nu_e$ prompt
neutronization burst~\cite{Duan:2007sh,Dasgupta:2008cd,Cherry:2010yc}. 

In conclusion, the time evolution of the SN neutrino signal provides 
a large amount of information on the flavor transformations of
neutrinos in the deepest stellar regions.
The early post-bounce deleptonization,  accretion and cooling
phases represent three different ``experiments''  that allow us to
probe different behaviors of the dense SN neutrino gas. 
In order to  learn the most from a future neutrino observation from a galactic
event, improved  theoretical and experimental 
studies are necessary to decipher the supernova neutrino enigma.

\section*{Acknowledgments} 

We thank B.~Dasgupta, T.~Janka, G.~Raffelt, C.~Ott and G.~Sigl for useful discussions and suggestions during the development of this project.
We also acknowledge E.~Lisi, G.~Raffelt, S.~Sarikas, and I.~Tamborra for helpful comments on the manuscript.
The work of S.C., A.M., N.S.  was supported by the German Science Foundation (DFG)
within the Collaborative Research Center 676 ``Particles, Strings and the
Early Universe''. T.F. acknowledges support from HIC for FAIR project~no.~62800075.

\section*{References} 

\end{document}